\documentclass[12pt,final]{article}
\usepackage{t1enc}
\usepackage[latin2]{inputenc}
\usepackage{showkeys}
\usepackage{graphicx}
\usepackage[margin=2cm]{geometry}
\usepackage{cite}
\usepackage{amssymb}
\begin{document}

\title{\Large \bf Effective Hamiltonian for fluid membranes in the presence of long-ranged forces}
\author{F. Dutka$^{1}$, M. Napi\'orkowski$^{1}$, and S. Dietrich$^{2,3}$\\ 
$^{1}$Instytut Fizyki Teoretycznej, Uniwersytet Warszawski\\  00-681 Warszawa, Ho\.za 69, Poland\\
$^{2}$Max-Planck-Institut f\"{u}r Metallforschung, Heisenbergstrasse 3, \\70569 Stuttgart, Germany \\
$^{3}$ Institut f\"{u}r Theoretische und Angewandte Physik, Pfaffenwaldring 57, \\
Universit\"{a}t Stuttgart, 70569 Stuttgart, Germany} 
\date{}
\maketitle{ }
\abstract{If the constituent particles of fluid phases interact via long-ranged van der Waals forces, the effective 
Hamiltonian for \emph{interfaces} between such fluid phases contains - in lateral Fourier space - non-analytic terms 
$ \sim q^4 \ln q$. Similar non-analytic terms characterize the effective Hamiltonian
for two interacting interfaces which can emerge between the three possible coexisting fluid phases in binary liquid mixtures. This 
is in contrast with the structure of the phenomenological Helfrich Hamiltonian 
for membranes which does not contain such non-analytic terms. We show that under favorable
conditions for the bulk densities characterizing a binary liquid mixture and for the long-ranged interparticle interactions 
the corresponding effective Hamiltonian for a model fluid \emph{membrane} does not exhibit such non-analytic 
contributions. We discuss the properties of the resulting effective Hamiltonian, with a particular emphasis on the influence 
of the long range of the interactions on the coefficient of the bending rigidity.}
\newpage
\section{Introduction}
In order to be able to describe nonplanar configurations of interfaces and membranes, the derivation and use of corresponding 
 effective Hamiltonians has been studied intensively \cite{Leibler1,Leibler2,Brochard}. Depending on the 
environment and their internal composition interfaces and membranes can display rather complex behaviors 
\cite{Seifert1}.  
A particular class of such systems is formed by the ubiquitous fluid-fluid interfaces and 
fluid membranes. In the case of interfaces the effective Hamiltonian takes on a capillary-wave like structure \cite{Evans1} 
while membranes are usually described in terms of the so-called 
Helfrich Hamiltonian \cite{Helf1}. \\
On the phenomenological level the effective Hamiltonian contains two types of contributions: the 
first is related to the possible change of the interface or membrane area and is controlled by the coefficient $\sigma_0$ of the surface tension 
while the second contribution is proportional to the square of the local mean curvature of the interface or membrane and is controlled 
by the coefficient $\kappa$ of the bending rigidity. In the following we consider fluctuating interfaces or membranes which are planar on the average 
and do not change their topology; thus contributions due to the Gaussian curvature do not matter. In lateral Fourier space the 
contribution from the $q$-mode $\tilde{f}({\bf q})$ of a local height configuration to the 
effective Hamiltonian is proportional to $|\tilde{f}({\bf q})|^2 \,(qa)^2\,\sigma(q)$, where $a$ is a microscopic length scale proportional 
to the particle diameter and $\sigma(q \to 0 )= \sigma_{0} \,+\, \kappa \, (qa)^2$. \\
Here we focus on the ensuing structure of 
the effective Hamiltonian for systems in which the interparticle interactions are of the long-ranged van der Waals type.  
This issue becomes acute if one tries to justify and to derive the phenomenological capillary-wave Hamiltonian from a microscopic 
theory such as, e.g., density 
functional theory \cite{Evans1}. In such approaches it turns out that for interfaces between fluid phases in systems governed by 
long-ranged forces the effective surface tension $\sigma(q)$ exhibits the form 
$\sigma(q \to 0) = \sigma_{0} \,+\, \sigma_{1} \, (qa)^2 \ln(qa) \, +\, \kappa \, (qa)^2$, and thus contains 
a leading non-analytic term  $\sigma_{1} \, (qa)^2 \ln(qa)$ with $\sigma_1>0$ which is not captured by phenomenological approaches.
This logarithmic singularity in Fourier space can be traced back to the divergence of the third and higher moments of the interparticle 
interaction potentials decaying as function of the distance $\sim r^{-6}$.  
For fluid interfaces the presence of such a non-analytic contribution has been established theoretically \cite{DN1,KN1,MD1,HDM1}. This 
implies that for small 
$q \neq 0$ one has $\sigma(q) < \sigma_0$ which has been confirmed also experimentally for various systems \cite{exp1} as well as in some simulations \cite{Milchev} but not in all \cite{Tarazona}. On the other 
hand such non-analytic terms are absent in the 
effective Helfrich Hamiltonian for membranes which, however, successfully describes various properties of fluid membranes. 
This is puzzling because the particles making up membranes invariably also exhibit long-ranged van der Waals interactions which in turn 
should lead to non-analytical bending contributions. \\
Our objective is to construct a simple model of a fluid membrane based on the extension of a model of two interacting 
fluid-fluid interfaces. We want to check under which conditions, if any, the absence of  non-analytic terms of the type 
$\sigma_{1} \, (qa)^2 \ln(qa)$ in the effective Hamiltonian for a membrane is possible, and  what kind of influence on the 
remaining terms these conditions have. In the following section we recall the relevant 
results concerning  the structure of the capillary-wave Hamiltonian.
In Sect. \ref{membrane} we discuss a simple model of fluid membranes in a system 
with long-ranged forces which is based on a model of two interacting fluid-fluid interfaces in a binary liquid mixture. We establish
the conditions under which the effective Hamiltonian for the fluid membrane 
is free from non-analyticities present in the corresponding capillary-wave Hamiltonian for the interface and  is thus compatible with the 
structure of the Helfrich Hamiltonian. In Sect. \ref{discussion} we compare our predictions for the resulting effective Hamiltonian with 
those discussed in the literature. 

\section{Effective Hamiltonian for a fluid-fluid interface}

In this section  we recall the basic facts pertinent to the structure of the capillary-wave Hamiltonian ${\cal H}_{cw}[f]$ 
for a fluid-fluid interface. Its local height relative to the reference plane $z=0$ is described by the function $z=f({\bf R})$, where 
${\bf R}=(x,y)$ denotes the lateral coordinates. Various aspects of this structure have been discussed in the literature. In particular, 
the issue of a local 
versus a non-local structure of ${\cal H}_{cw}[f]$ has been extensively analyzed for the cases of short-ranged (exponentially) 
and long-ranged (algebraicly) decaying interactions \cite{DN1,KN1,MD1,HDM1,Par1}. A suitable framework for analyzing such issues is density 
functional theory for non-uniform fluids. This analysis is particularly straightforward if 
the non-uniform one-component fluid 
density $\rho({\bf R},z)$ associated with an interface configuration $f({\bf R})$ is approximated within the so-called sharp-kink 
approximation by a piecewise constant function 
$\rho_{shk}({\bf R},z) = \rho_{\beta}\,\Theta(f({\bf R})-z)\, + \, \rho_{\alpha}\,\Theta(z-f({\bf R}))$, 
where $\rho_{\alpha}$ and $\rho_{\beta}$ denote the bulk densities of the coexisting fluid phases $\alpha$ and $\beta$, and $\Theta(z)$ 
denotes the Heaviside function. If, moreover, the effective Hamiltonian is truncated to be bilinear in $f$, it can be written as \cite{DN1,KN1,MD1} 
\begin{eqnarray}
\label{efect1} 
{\cal H}_{cw}[f] = \frac{1}{2} \int \frac{d^2 q}{(2 \pi)^2} \, \left|\tilde{f}({\bf q})\right|^2 \, q^2 \, \sigma(q) 
\end{eqnarray}
where 
\begin{eqnarray} 
\tilde{f}({\bf q}) = \int d^2 R \, f({\bf R}) \, \exp(-i {\bf q}\cdot {\bf R}) \quad.  
\end{eqnarray}
The wavevector dependent surface tension $\sigma(q)$ in Eq.(\ref{efect1}) is given by   
\begin{eqnarray} 
\sigma(q) = q^{-2}\, [\,\tilde{w}(q) - \tilde{w}(0)\,] \left(\rho_{\alpha} - \rho_{\beta}\right)^2 \quad,
\end{eqnarray}
where $\tilde{w}(q)$ denotes the Fourier transform of the long-ranged part of the spherically symmetric interparticle interaction potential 
$w(r=|({\bf R},z)|)$ 
taken with respect to the lateral coordinates for $z=0$:  
\begin{eqnarray} 
\tilde{w}(q) = \int d^2 R \, w(|({\bf R},z=0)|) \, \exp(-i {\bf q}\cdot {\bf R}) \quad.  
\end{eqnarray}
It has turned out to be suitable to adopt for the long-ranged part of the van der Waals pair potential $w({\bf R},z)$ the form 
\begin{eqnarray} 
w(|({\bf R},z)|) = - \, \frac{A}{({\bf R}^2+z^2+a^2)^3} \quad,
\end{eqnarray}
where $a$ corresponds to the hard core radius of the fluid particles and $A>0$ characterizes the strength of 
the attractive interparticle interaction. For  $\bar{q} = qa \ll 1$ the ensuing $\sigma(q)$ has  the following 
non-analytic form:
\begin{eqnarray}
\label{nonan1} 
\sigma(q) \, = \, \sigma_{0} + \sigma_{1} \, \bar{q}^2\,\ln(\bar{q}) + \sigma_{2} \, \bar{q}^2 +O(\bar{q}^4) \quad, 
\end{eqnarray}
where $\sigma_{0}\,=\, \frac{A\,\pi}{8\,a^2}\,(\rho_{\alpha} - \rho_{\beta})^2  > 0$, $\sigma_{1}\,=\,\frac{1}{4} \sigma_{0} > 0 $, 
$\sigma_{2}\,=\,\frac{1}{4}\,\sigma_{0}\,C_{0}<0$, $C_{0}=C_{\cal{E}}-3/4-\ln2 = -0.866$, and 
$C_{\cal{E}}$ denotes Euler's constant.  \\
A more realistic approach to determine $\sigma(q)$ \cite{MD1} takes into account the influence of local interfacial curvatures on 
the actual smooth intrinsic density profile. The effective Hamiltonians for interfaces both in one-component \cite{MD1} and in binary 
liquid mixtures \cite{HDM1} have been analyzed along these lines. For long-ranged van der Waals interactions in each case the 
presence of non-analytic terms in $\sigma(q)$ (Eq.(\ref{efect1})) has been established.

\section{A model of a fluid membrane \label{membrane}}

For the comparison between effective Hamiltonians 
for fluid-fluid interfaces and fluid membranes it is particularly suitable to consider binary liquid mixtures. Upon special choices of the thermodynamic conditions these systems 
allow  for the coexistence of three fluid phases denoted as 
$\alpha$, $\beta$, and $\gamma$. In the presence of appropriately chosen boundary conditions or external ordering fields one can 
consider a situation in which a layer of - say - phase $\beta$ with mean thickness $\ell$ separates the phases $\alpha$ and $\gamma$ \cite{Getta}. 
In such a system there are two fluid-fluid interfaces the positions of which are denoted by 
$f_{\alpha\beta}({\bf R})$ and $\ell + f_{\beta\gamma}({\bf R})$. They  separate the phases $\alpha$, $\beta$ and $\beta$,$\gamma$, 
respectively, 
(see Fig. 1).   
\begin{figure}[htb]
 \centering
  \includegraphics[width = 0.5 \textwidth]{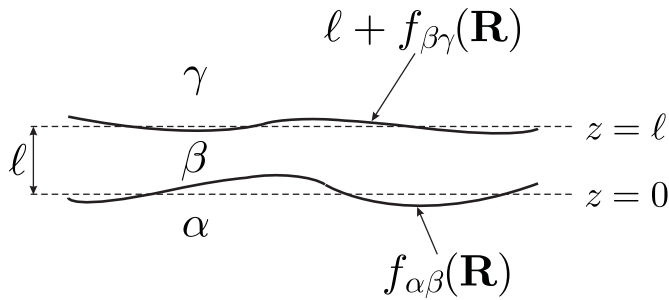}
  \caption{The system under consideration consists of two fluid-fluid interfaces 
$f_{\alpha\beta}({\bf R})$ and $\ell + f_{\beta\gamma}({\bf R})$ separating the phases $\alpha$, $\beta$ and $\beta$, $\gamma$,  respectively,
and fluctuating around their mean positions $z=0$ and $z=\ell$,  respectively.}
\end{figure}
We note that although the system is characterized by six number densities $\rho_{i\lambda}$, with $i=1,2$, 
and $\lambda=\alpha, \beta, \gamma$, where $\rho_{i\lambda}$ denotes  the number density of the $i$-th component 
in phase $\lambda$, three-phase coexistence allows for only one independent
thermodynamic variable such as temperature; on the corresponding triple line the chemical potentials $\mu_1(T)$ and $\mu_2(T)$ of the two species are fixed. 
In addition there are three interparticle interactions present in the system: two among the two species and one between the different species. 
They are assumed to be spherically symmetric 
and are denoted by $w_{ij}( r)=w_{ji}(r)=\,w_{ij}(|({\bf R},z)|)$ with $i,j=1,2$. \\ 
 
For such a system containing two interfaces the capillary-wave Hamiltonian \linebreak ${\cal H}_{cw}([f_{\alpha\beta}, f_{\beta\gamma}], \ell)$ 
is a functional 
of the two interfacial positions $f_{\alpha\beta}({\bf R})$ and $f_{\beta\gamma}({\bf R})$ and a function of the distance $\ell$. 
Applying the sharp-kink approximation to the density functional for binary liquid mixtures described, e.g., in Refs. \cite{KN1,HDM1} yields within
the bilinear approximation, which retains non-locality, the following form:
\begin{eqnarray} 
{\cal H}_{cw}([f_{\alpha\beta}, f_{\beta\gamma}], \ell) = \frac{1}{2}\,\int \frac{d^2 q}{(2\pi)^2}\,
\left\{ \, 2 \,\tilde {f}_{\alpha\beta}({\bf q}) \,\tilde {f}_{\beta\gamma}(-{\bf q}) \,\,\tilde{w}_{\alpha\beta, \beta\gamma}(q,\ell) \right.\\
\left. + \, \left|\tilde {f}_{\alpha\beta}({\bf q}) \right|^2 \left[q^2\,\sigma_{\alpha\beta}(q) \,-\,
\tilde{w}_{\alpha\beta, \beta\gamma}(0,\ell)\,\right] 
+\, \left|\tilde {f}_{\beta\gamma}({\bf q})\right|^2\,\left[q^2\,\sigma_{\beta\gamma}(q) \, - \,
\tilde {w}_{\alpha\beta, \beta\gamma}(0,\ell)\,\right] \right\} \ , \nonumber
\end{eqnarray}
where 
\begin{eqnarray}
 \tilde{w}_{\lambda\kappa, \eta\delta}(q,z)\,=\, \sum_{i,j=1}^{2}(\rho_{i\lambda}-\rho_{i\kappa})(\rho_{j\eta}-\rho_{j\delta})
\int d^2 R\,\exp(-i{\bf q}\cdot{\bf R})\, w_{ij}({\bf R},z) 
\end{eqnarray}
and \cite{KN1,HDM1}
\begin{eqnarray}
\sigma_{\lambda \kappa}(q)=q^{-2} \left[\tilde{w}_{\lambda\kappa, \lambda\kappa}(q,z=0)-\tilde{w}_{\lambda\kappa, \lambda\kappa}(q=0,z=0)\right] \ .
\end{eqnarray}
The above results can serve as a starting point to construct a simple model of a fluid membrane. To this end we take the 
two interface configurations to be in phase, i.e., $f_{\alpha\beta}({\bf R}) = f_{\beta\gamma}({\bf R})$. This renders a model fluid 
membrane consisting of phase $\beta$ embedded on one side by phase $\alpha$ and on the other side by phase $\gamma$. 
The thickness $\ell$ of the membrane is uniform and its upper and lower boundaries have the same shape described by 
$f({\bf R})\,=\, f_{\alpha\beta}({\bf R}) = f_{\beta\gamma}({\bf R})$. In this case and within the bilinear 
approximation the capillary-wave Hamiltonian reduces to:
\begin{eqnarray}
\label{efectiveH}
{\cal H}([f], \ell) = \frac{1}{2}\,\int \frac{d^2 q}{(2\pi)^2}\,\left|\tilde {f}({\bf q}) \right|^2 \,\gamma(q) \, q^2 \quad,
\end{eqnarray} 
with 
\begin{eqnarray}
q^2\,\gamma(q)\,=\,\sum_{i,j=1}^{2}\left\{\left[\,(\rho_{i\alpha}-\rho_{i\beta})(\rho_{j\alpha}-\rho_{j\beta})\,+\,
(\rho_{i\beta}-\rho_{i\gamma})(\rho_{j\beta}-\rho_{j\gamma})\right][\tilde{w}_{ij}(q,0)-\tilde{w}_{ij}(0,0)] \right.\,\nonumber \\
\left.+\,2\,(\rho_{i\alpha}-\rho_{i\beta})(\rho_{j\beta}-\rho_{j\gamma})[\tilde{w}_{ij}(q,\ell)-\tilde{w}_{ij}(0,\ell)]\,\right\} \ .
\end{eqnarray}
With the choice    
\begin{eqnarray} 
w_{ij}(|({\bf R},z)|) = - \, \frac{A_{ij}}{({\bf R}^2+z^2+a_{ij}^2)^3}
\end{eqnarray}
for the long-ranged interparticle potentials one has 
\begin{eqnarray} 
\tilde{w}_{ij}(q,\ell)-\tilde{w}_{ij}(0,\ell) \,=\,\frac{\pi}{8}\,A_{ij}\,\left[\frac{q^2}{a_{ij}^2\,+\,\ell^2}\,+\,
\frac{q^4}{4}\,\left(\ln(q\sqrt{a_{ij}^2\,+\,\ell^2})\,+\,C_{0}\right)\right] \ .
\end{eqnarray}
For reasons of simplicity in the following we assume $a_{ij}=a, \, i,j=1,2$. This choice leads to the following expression for $\gamma(q)$:
\begin{eqnarray}
\label{gammaq}
\gamma(q)\,=\,\gamma_{0} \,+\,\gamma_{1}\,\bar{q}^2\,\ln\bar{q}\,+\,\gamma_{2}\,\bar{q}^2 + O(\bar{q}^4) \ ,
\end{eqnarray}
where, with $\bar \ell = \ell/a$,
\begin{eqnarray}
\gamma_{0}(\ell) \,=\,\frac{\pi}{8\,a^2}\,\sum_{i,j=1}^{2}\,A_{ij}\,\left\{ \,\left[(\rho_{i\alpha}-\rho_{i\beta})(\rho_{j\alpha}-\rho_{j\beta})\,
+\,(\rho_{i\beta}-\rho_{i\gamma})(\rho_{j\beta}-\rho_{j\gamma})\right] \,\right. \nonumber \\
\left. + \frac{2}{1+\left. \bar{\ell}\right.^2}
(\rho_{i\alpha}-\rho_{i\beta})(\rho_{j\beta}-\rho_{j\gamma})\,\right\} \ ,
\end{eqnarray}
\begin{eqnarray}
\label{gamma1}
\gamma_{1}\,=\,\frac{\pi}{32\,a^2}\,\sum_{i,j=1}^{2}\,A_{ij}\,(\rho_{i\alpha}-\rho_{i\gamma})(\rho_{j\alpha}-\rho_{j\gamma}) \quad,
\end{eqnarray}
and
\begin{eqnarray}
\gamma_{2}(\ell)\,=\,\frac{\pi}{32\,a^2}\,\sum_{i,j=1}^{2}\,A_{ij}\,\left\{\ln\left(1+\left.\bar{\ell}\right.^2\right)\,
(\rho_{i\alpha}-\rho_{i\beta})(\rho_{j\beta}-\rho_{j\gamma})\, \right. \nonumber \\
\left. +\,C_{0}\,(\rho_{i\alpha}-\rho_{i\gamma})(\rho_{j\alpha}-\rho_{j\gamma})\right\} \ .
\label{gamma2}
\end{eqnarray}
As expected, similar to the case of single interface (Eq.(\ref{nonan1})) the effective Hamiltonian for the model fluid membrane contains 
a non-analytic contribution $\gamma_{1}\,\bar{q}^4\,\ln\bar{q}$. In this sense the structure of the effective Hamiltonian 
given by Eqs.(\ref{efectiveH},\ref{gammaq}-\ref{gamma2}) is not compatible 
with the phenomenological Helfrich Hamiltonian ansatz which for small membrane ondulations can be expressed in its form as in 
Eq.(\ref{gammaq}) but with $\gamma_{1}=0$. \\

Our purpose is thus to find conditions under which the coefficient $\gamma_{1}$ of the non-analytic contribution in 
Eq.(\ref{gammaq}) vanishes. There are two particularly simple choices of the number densities 
$\rho_{i\lambda}$ and the amplitudes $A_{ij}$ of the interaction potentials which fulfill this requirement. 
The first choice $(I)$ puts constraints on the densities of the phases $\alpha$ and $\gamma$ and stipulates 
\begin{eqnarray}
\label{cond1}
 \rho_{i\alpha}\,=\,\rho_{i\gamma} \qquad \qquad i=1,2 \qquad (I) \ . 
\end{eqnarray}
This condition imposes that the two phases on both sides of the membrane are identical. The second choice 
($II$) puts constraints both on the interaction amplitudes and on the densities. First, it requires that   
\begin{eqnarray}
\label{cond21}
 A_{12}\,=\,\sqrt{A_{11}\,A_{22}} \qquad (IIa)
\end{eqnarray}
which leads to 
\begin{eqnarray}
\label{gamma11}
\gamma_{1}\,=\,\frac{\pi}{32\,a^2} \left[ \sqrt{A_{11}}\,(\rho_{1\alpha}-\rho_{1\gamma})\,+\, 
\sqrt{A_{22}}\,(\rho_{2\alpha}-\rho_{2\gamma}) \right]^2 \ .
\end{eqnarray}
The additional requirement  
\begin{eqnarray}
\label{cond22}
 \sqrt{A_{11}}\,(\rho_{1\alpha}-\rho_{1\gamma})\,=\, -\sqrt{A_{22}}\,(\rho_{2\alpha}-\rho_{2\gamma}) \qquad (IIb)
\end{eqnarray}
implies $\gamma_{1}=0$.
It is straightforward to show that the above condition $I$ (Eq.(\ref{cond1})) leads to 
\begin{eqnarray}
\label{gammaI}
\gamma^{(I)}(q)\,=\,\frac{\pi}{4\,a^2}\,\left[\sqrt{A_{11}}(\rho_{1\alpha}-\rho_{1\beta})\,+\,
\sqrt{A_{22}}(\rho_{2\alpha}-\rho_{2\beta})\,\right]^2\,\varphi\left(\bar{q},\bar{\ell}\right) 
\end{eqnarray}
where
\begin{eqnarray}
\label{varphi1}
\varphi\left(\bar{q},\bar{\ell}\right)\,=
\,\frac{\left.{\bar \ell}\right.^2}{1\,+\,\left.\bar{\ell}\right.^2}\,-\,\frac{\bar{q}^2}{8}\ln\left(1\,+\,\left.\bar{\ell}\right.^2\right)  \ .
\end{eqnarray}
Interestingly, if the conditions $(IIa)$ (Eq.(\ref{cond21})) and $(IIb)$ (Eq.(\ref{cond22})) are imposed, 
the corresponding effective surface tension $\gamma^{(II)}(q)$ has exactly the same form as for the first condition, i.e., 
$\gamma^{(II)}(q)\,=\,\gamma^{(I)}(q)$. 
The fact that the requirements $(II)$, which put constraints on both the densities and the interaction amplitudes, lead to 
the same result as the requirement $(I)$, which identifies the phases $\alpha$ and $\gamma$ but does not involve the interaction 
amplitudes $A_{ij}$, can be understood as follows. We consider a typical contribution to the free-energy density functional which describes 
the interaction between particles located in a 
region $\cal{V}_{\alpha}$ of the binary liquid mixture with a specific particle of type $k$, $k=1,2$, located at ${\bf r}'$ 
somewhere in the system. This term is proportional to 
\begin{eqnarray}
\int_{\cal{V}_{\alpha}} d^3 r  \sum_{i} w_{ik}({\bf r} - {\bf r}') \rho_{i \alpha} \sim
\sum_{i} A_{ik} \rho_{i \alpha} = A_{1k} \rho_{1 \alpha} +  A_{2k} \rho_{2 \alpha}  \qquad \qquad \qquad \qquad \qquad \qquad \quad   \\
=  \sqrt{A_{kk}} \left(\sqrt{A_{11}} \rho_{1 \alpha} + \sqrt{A_{22}} \rho_{2 \alpha} \right) 
= \sqrt{A_{kk}} \left(\sqrt{A_{11}} \rho_{1 \gamma} + \sqrt{A_{22}} \rho_{2 \gamma} \right) 
= A_{1k} \rho_{1 \gamma} +  A_{2k} \rho_{2 \gamma}  \ , \nonumber
\end{eqnarray}
where the conditions in Eq.(\ref{cond21}) and Eq.(\ref{cond22}) have been used. One concludes that this 
contribution to the free-energy functional has the same form as if the region $\cal{V}_{\alpha}$ would be  
filled with particles with densities $\rho_{i \gamma}$ instead of $\rho_{i \alpha}$. But this is 
exactly the requirement in Eq.(\ref{cond1}) corresponding to choice $(I)$ which identifies the phases 
$\alpha$ and $\gamma$.  \\
In the next section we discuss the properties of the resulting effective Hamiltonian. 

\section{Discussion \label{discussion}}

In the previous section we showed that for special choices for the densities or  
the interparticle interactions in binary liquid mixtures there are no non-analytic contributions to $\gamma(q)$ in the limit of small $q$ (up to and including $O(q^2)$. 
This choice eliminates the leading non-analytic contribution for any membrane thickness $\ell$, because $\gamma_1$ does not 
depend on $\ell$ (see Eq.(\ref{gamma1})).
It turns out that independent of  whether constraints of type $(I)$ in Eq.(\ref{cond1}) or of type $(II)$ 
in Eqs.(\ref{cond21}, \ref{cond22}) are imposed the resulting effective Hamiltonian for the model fluid membrane takes the form 
given by Eqs.(\ref{efectiveH}) and (\ref{gammaI}). The function $\gamma^{(I)}(q)$ in  Eq.(\ref{gammaI}) is determined by  
the bulk number densities $\rho_{i\lambda},  i=1,2, \lambda=\alpha, \beta$, the interaction strengths $A_{11}, A_{22}$, the particles 
size $a$, and the membrane thickness $\ell$. The function $\gamma^{(I)}(q)$ factorizes into a product of two functions. The first 
factor depends on the densities and interaction strengths only and is non-negative. The second factor depends on $\bar{q}$ and 
parametrically on $\bar{\ell}$ only; the parameter $a$ sets the scale for the variables $q$ and $\ell$.   This second factor, which we
denoted as $\varphi(\bar{q}, \bar{\ell})$, is 
particularly interesting because - contrary to the first factor - it can change sign depending on 
the  values of $\bar{q}$ and $\bar{\ell}$. This possibility of $\gamma^{(I)}(q)$ to change sign appears because the 
coefficient $\gamma_{2}$ in Eq.(\ref{gammaq}) is inherently negative, i.e., the contribution from the long-ranged forces to 
the coefficient of the bending rigidity is negative. (Note that within the sharp-kink approximation, which takes only the influence of the long-ranged forces into account, also $\sigma_2$ is negative (see Eq.(\ref{nonan1}) and the expressions following it).)
This conclusion checks qualitatively 
with a recent analysis by Dean and Horgan \cite{DH1} who have expressed the coefficient of the bending rigidity in terms of the membrane 
thickness and the dielectric constants of the membrane ($\epsilon$) and of the surrounding medium ($\epsilon'$):
\begin{eqnarray}
\gamma^{DH}_{2}(\ell)\,=\,-\,\frac{3k_{B}T}{128\pi}\,\left(\frac{\epsilon-\epsilon'}{\epsilon+\epsilon'}\right)^2
	\,\ln\left(1+\left.\bar{\ell}\right.^2\right) \ .
\end{eqnarray}
Dean and Horgan \cite{DH1} have not discussed the issue of the presence of non-analytic terms in the effective Hamiltonian. However,   
our result and those in \cite{DH1} agree concerning the functional form of the dependence of the coefficient $\gamma_{2}(\ell)$  
on the membrane thickness $\ell$. 
Within both approaches the coefficient of the bending rigidity depends logarithmically on the membrane thickness, i.e., 
$\gamma_{2} \sim \ln(1+\left.\bar{\ell}\right.^2)$. Of course realistic  membrane models yield additional contributions to the bending rigidity 
stemming from other types of interactions present in the system. In our approach only the long-ranged contributions to the 
bending rigidity are considered. In this latter case the negative coefficient $\gamma_2$ of the bending rigidity in the 
presence of the positive coefficient $\gamma_0$ of the surface tension leads to an instability at small wavelengths of 
the membrane ondulations. According to Eq.(\ref{varphi1}) this instability occurs for 
\begin{eqnarray}
\label{unst1}
\bar{q}^2 > \frac{8\,\left.\bar{\ell}\right.^2}{(1\,+\,\left.\bar{\ell}\right.^2)\,\ln(1\,+\,\left.\bar{\ell}\right.^2)} \qquad (unstable; I, II) \ .  
\end{eqnarray}
On the other hand the wavevectors must be smaller than the physically allowed maximal one $  \bar q_{max} \lesssim 1$.
This implies that the values of $\bar \ell$ for which the instability can occur fulfill the condition
\begin{eqnarray}
\bar{\ell} \, > \,\bar \ell_0 = \exp\left(\frac{4}{\bar{q}_{max}^{2}}\right) \qquad (unstable; I, II) \ . 
\end{eqnarray}
For $\bar{q}_{max}\,=\,1/2$ one has $\bar{\ell}_0 \, =  9 \times 10^6$. This condition states that for membrane thicknesses 
$0 < \bar{\ell} < \bar \ell_0 $ the negative bending rigidity coefficient does not give rise to instabilities for 
membrane ondulations with wavevectors within the physically accessible range $\bar{q} < \bar{q}_{max}$.  \\

Finally we mention that the vanishing of the coefficient $\gamma_{1}$ can also occur in binary liquid 
mixtures in which the interactions $w_{11}$ and $w_{22}$ are repulsive, i.e., $A_{11}, A_{22} < 0$  while the  
interactions $w_{12}$ are attractive, i.e., $A_{12} > 0$. (It is conceivable that such a situation may arise in multicomponent complex 
fluids with effective interactions between two dominating species upon integrating out the degrees of freedom of the smaller species. This can occur if the two species are oppositely charged.)  
This is a different situation from the one considered above in which all long-ranged interactions were assumed to 
be attractive, i.e., $A_{11}, A_{22}, A_{12} > 0$. In this present case the conditions $(IIa)$ and $(IIb)$ are replaced by 
\begin{eqnarray}
 A_{12}\,=\,\sqrt{(-A_{11})\,(-A_{22})} \qquad (IIIa)
\end{eqnarray}
so that 
\begin{eqnarray}
\gamma_{1}\,=\,-\,\frac{\pi}{32\,a^2}\, \left[\sqrt{-A_{11}}\,(\rho_{1\alpha}-\rho_{1\gamma})\,-\, 
\sqrt{-A_{22}}\,(\rho_{2\alpha}-\rho_{2\gamma})\right]^2 \ ,
\end{eqnarray}
and by
\begin{eqnarray}
 \sqrt{-A_{11}}\,(\rho_{1\alpha}-\rho_{1\gamma})\,=\, \sqrt{-A_{22}}\,(\rho_{2\alpha}-\rho_{2\gamma}) \qquad (IIIb) \ ,
\end{eqnarray}
respectively. 
It is straightforward to see that in this case the effective surface tension denoted as $\gamma^{(III)}(q)$ is given by  
\begin{eqnarray}
\gamma^{(III)}(q)\,=\,-\frac{\pi}{4\,a^2}\,\left[\sqrt{-A_{11}}(\rho_{1\alpha}-\rho_{1\beta})\,-\,
\sqrt{-A_{22}}(\rho_{2\alpha}-\rho_{2\beta})\,\right]^2\,\varphi\left(\bar{q},\bar{\ell}\right) \ .
\end{eqnarray}
Accordingly the model fluid membrane is unstable with respect to long-wavelength ondulations:
\begin{eqnarray}
\label{unst2}
\bar{q}^2 < \frac{8\,\left.\bar{\ell}\right.^2}{(1\,+\,\left.\bar{\ell}\right.^2)\,\ln(1\,+\,\left.\bar{\ell}\right.^2)} \qquad (unstable; III) \ .  
\end{eqnarray}
This implies that for ondulations with given $\bar{q}$-values only membranes with thicknesses 
\begin{eqnarray}
\bar{\ell} \, < \, \exp\left(\frac{4}{\bar{q}^{2}}\right) \qquad (unstable; III)
\end{eqnarray}
are unstable. \\
To summarize, we have shown that it is possible to choose conditions under which the leading non-analytic contribution  
to the effective Hamiltonian of a fluid membrane in the presence of long-ranged forces vanishes. One of them amounts 
to the requirement that the embedding phases on both sides of a fluid membrane are identical. We have also checked that the 
contribution from long-ranged forces to the coefficient of the bending rigidity is negative and we have discussed the  
implications on the stability of membranes with respect to ondulations. \\

\vspace{2cm}  
{\large \bf Acknowledgements:}

The work of F.D. and M.N. has been financed from funds provided for scientific research for the years 2006-2008 under the research project N202 076 31/0108.

\end{document}